\documentclass[conference]{IEEEtran}

\usepackage{cite}

\usepackage{amsmath}
\usepackage{mathtools}
\usepackage{amssymb}
\usepackage{amsfonts}

\usepackage{array}
\usepackage{multirow}
\usepackage{dcolumn}
\usepackage{booktabs}

\DeclareMathOperator{\ima}{im}

\usepackage{graphicx}
\usepackage{epstopdf}
\usepackage{float}
\usepackage{subfloat}
\usepackage{subfig}

\usepackage{algorithmic}
\usepackage{algorithm}

\usepackage{xspace}

\makeatletter
\newcommand*{\etc}{%
    \@ifnextchar{.}%
        {etc}%
        {etc.\@\xspace}%
}
\makeatother

\usepackage{tikz}
\usetikzlibrary{shapes,arrows,automata}
\usetikzlibrary{decorations.markings}

\begin{document}



\title{Distributed Computation of the \v{C}ech Complex\\and Applications in Wireless Networks}

\author{\IEEEauthorblockN{Ngoc-Khuyen LE, Ana\"is Vergne, Philippe Martins, Laurent Decreusefond} 
\IEEEauthorblockA{LTCI, CNRS, T\'el\'ecom ParisTech, Universit\'e Paris-Saclay, 75013, Paris, France\\
Email: \{ngoc-khuyen.le, anais.vergne, philippe.martins, laurent.decreusefond\}@telecom-paristech.fr
}
}


%


\maketitle

\begin{abstract}
In this paper, we introduce a distributed algorithm to compute the \v{C}ech complex. This algorithm is aimed at solving coverage problems in self organized wireless networks. Two applications based on the distributed computation of the \v{C}ech complex are proposed. The first application detects coverage holes while the later one optimizes coverage of wireless networks.
\end{abstract}

\section{Introduction}

Coverage is a key factor that determines the quality of service in wireless networks. In recent applications to solve coverage problems in wireless networks, simplicial complex is often utilized to represent the network topology. Unlike graph, which represents only the neighborhood of cells, simplicial complex presents the relationship of cells with a higher dimension. 

The \v{C}ech complex is a simplicial complex that represents exactly the topology of the network~\cite{Silva-RSS-05}. The \v{C}ech complex is the right tool to describe and optimize the coverage for wireless networks. However, the computation of the \v{C}ech complex is complicated. The \v{C}ech complex represents a collection of cells by a simplex if they have a non-empty intersection. In~\cite{Dantchev2012708}, an algorithm to compute the \v{C}ech complex is introduced, but this algorithm is designed to be utilized in graphics science. Therefore, this algorithm only works with a collection of same sized cells. Although an algorithm to compute the \v{C}ech complex for a collection of differently sized cells is proposed in~\cite{7145759}, this algorithm is still centralized. 

In~\cite{7247173}, authors introduce an application based on the \v{C}ech complex to optimize the coverage for wireless networks. However, as the computation of the \v{C}ech complex is still available in centralized way, this application is also centralized. The future wireless networks will be self organized and therefore require distributed computations. In~\cite{Ghrist:2005:CHS:1147685.1147729, DeSilva:2006:CCS:1274636.1274646, 6134128}, some algorithms to detect coverage holes in wireless networks are developed. Although these algorithms are available in both centralized and distributed way, they use the Rips complex instead of the \v{C}ech complex to represent the topology of the network. The Rips complex is an approximation of the \v{C}ech complex. The Rips complex represents a collection of cells by a simplex if every pair of cells in this collection are intersected. Therefore, the Rips complex may not capture exactly the topology of the network. So, some coverage holes may be undiscovered~\cite{6364341}. 

In this paper, we introduce the distributed computation of the \v{C}ech complex for a given collection of differently sized cells. This algorithm is aimed at solving coverage problems in wireless networks. We also propose two applications that are based on the distributed computation of the \v{C}ech complex. Given a wireless network, the first application detects coverage holes while the second one optimizes its coverage. The optimized coverage reduces as much as possible the intersection among cells. As a result, the waste power due to interference within intersected regions is avoided efficiently. Both of these applications are distributed.

This paper is organized as follows. Section~\ref{sec:SimplicialHomology} introduces a short background of simplicial homology and its  applications in wireless networks. Section~\ref{sec:dist_computation_Cech_complex} describes all details about the distributed computation of the \v{C}ech complex. Section~\ref{sec:dist_Cech_complex_application} proposes a distributed coverage hole detection and a distributed coverage optimization for wireless networks. The last section concludes our paper.

\section{Simplicial homology and applications}
\label{sec:SimplicialHomology}

In this section, we introduce some notions of simplicial homology and its applications. For further details about the simplicial homology, see documents \cite{MunkresJ1984} and \cite{AllenHatcher}.  

\begin{figure}[h!]
\centering
\begin{tikzpicture}[scale = 0.40]
\begin{scope}[shift={(-1,0)}]
		\filldraw	[black,thick]	(-3.5,-0.5)node[above]{$v_0$}circle(2pt)	(-3.5,-1.75)node[below]{\(0\)-simplex};
		\draw	[black,thick]	(0.25,-0.75) -- (1.75,0.0)		(1,-1.75)node[below]{\(1\)-simplex} (0.25,-0.75)node[left]{$v_0$} (1.75,0.0)node[right]{$v_1$};
		\filldraw	[black,thick]	(0.25,-0.75)	circle(2pt);
		\filldraw	[black,thick]	(1.75,0.0)	circle(2pt);
\end{scope}
\begin{scope}[shift={(1,0)}]
		\filldraw	[blue!25]	(3,-0.75) -- (5,-0.75) -- (4,1);
		\draw	[black,thick]	(3,-0.75) -- (5,-0.75) -- (4,1) -- cycle;	
		\draw (4,-1.75)node[below]{\(2\)-simplex};
		\filldraw	[black,thick]	(3,-0.75)circle(2pt)node[left]{$v_0$}	(5,-0.75)circle(2pt)node[right]{$v_1$}	(4,1)circle(2pt)node[above]{$v_2$};
\end{scope}
\begin{scope}[shift={(3.2,0)}]
		\filldraw	[blue!50]	(6,-0.25) -- (7.5,-0.75) -- (8.5,0.5) -- (6.5,1);
		\draw	[black,thick]	(6,-0.25) -- (7.5,-0.75) -- (8.5,0.5) -- (6.5,1) -- cycle;	
		\draw	[black,thick]	(6.5,1) -- (7.5,-0.75)	;	
		\draw	[black,thick,dashed]	(6,-0.25) -- (8.5,0.5);	
		\draw	(7.25,-1.75)node[below]{\(3\)-simplex};
		\filldraw	[black,thick]	(6,-0.25)circle(2pt)node[left]{$v_0$} (7.5,-0.75)circle(2pt)node[below]{$v_1$} (8.5,0.5)circle(2pt)node[right]{$v_2$} (6.5,1)circle(2pt)node[above]{$v_3$};
\end{scope}
\end{tikzpicture}
\caption{An example of simplices.}
\label{fig:example_simplices}
\end{figure}
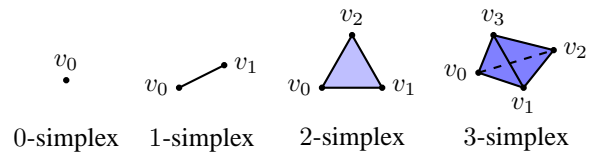

Given a set of vertices $V$, a $k$-simplex is an unordered subset $\{v_0, v_1, \ldots, v_k\}$, where $v_i \in V$ and $v_i \neq v_j$ for all $i \neq j$. The number $k$ is its dimension. See Figure~\ref{fig:example_simplices} for some instances. Any subset of vertices set of a simplex is a face of this simplex.
An oriented simplex is an ordered type of simplex. The orientation of a simplex changes if two vertices of this simplex swap their position. The change of orientation is represented by a negative sign as:
\begin{equation*}
[v_0, v_1, \ldots, v_i, v_j, \ldots, v_k] = - [v_0, v_1, \ldots, v_j, v_i, \ldots, v_k]
\end{equation*}

An abstract simplicial complex is a collection of simplices such that: every face of a simplex is also in the simplicial complex.

Let $X$ be a simplicial complex. For each $k \geq 0$, we define a vector space $C_k(X)$ whose basis is a set of oriented $k$-simplices of $X$. If $k$ is bigger than the highest dimension of $X$, let $C_k(X) = 0$. We define the boundary operator to be a linear map $\partial : C_k \to
C_{k-1}$ as follows:
\begin{equation*}
\partial[v_0, v_1, \ldots, v_k] = \sum_{i = 0}^{k} (-1)^{i}[v_0, v_1, \ldots, v_{i-1}, v_{i+1}, \ldots, v_k]
\end{equation*}

\begin{figure}[h!]
\centering
	\begin{tikzpicture}[scale=0.5, decoration={%
   markings,%
   mark=at position 0.5 with {\arrow[black]{stealth};},%
   }]
	\begin{scope}[shift = {(-1.5,0)}]
		\draw	[postaction=decorate] [black,thick]	(0,0)node[left]{$v_0$} -- (2,0)node[right]{$v_1$};
		\filldraw	[black,thick]	(0,0)	circle(2pt);
		\filldraw	[black,thick]	(2,0)	circle(2pt);
	\end{scope}
	\begin{scope}[shift = {(-1,0)}]
		\draw [black,thick]	(3.5,0)node{$\xrightarrow{\partial}$};	
	\end{scope}
	\begin{scope}[shift = {(0,0)}]
		\filldraw [black,thick]	(5,0)circle(2pt)node[left]{$v_0 +$};
		\filldraw [black,thick]	(7,0)circle(2pt)node[right]{$v_1 -$};
	\end{scope}
	\begin{scope}[shift = {(0.5,0)}]
			\draw [black,thick]	(9,0)node{$\xrightarrow{\partial}$};
	\end{scope}

	\begin{scope}[shift = {(0,0)}]
		\draw (11,0)node{$0$};
	\end{scope}
	\begin{scope}[shift = {(-2,0)}]
		\filldraw	[blue!25]	(0,-3) -- (2,-3) -- (1,-1.25);
		\draw	[black,thick]	(0,-3)node[left]{$v_0$} -- (2,-3)node[right]{$v_1$} -- (1,-1.25)node[above]{$v_2$} -- cycle;
		\filldraw	[black,thick]	(0,-3)circle(2pt) -- (2,-3)circle(2pt) -- (1,-1.25)circle(2pt);
	\end{scope}
	\begin{scope}[shift = {(-1,0)}]
			\draw [black,thick]	(3.5,-2)node{$\xrightarrow{\partial}$};
	\end{scope}
	\begin{scope}[shift = {(0,0)}]
		\draw	[postaction=decorate] [black,thick]	(5,-3)node[left]{$v_0$} -- (7,-3)node[right]{$v_1$};
		\draw	[postaction=decorate] [black,thick] (7,-3)node[right]{$v_1$} -- (6,-1.25)node[above]{$v_2$};
		\draw	[postaction=decorate] [black,thick] (6,-1.25)node[above]{$v_2$} -- (5,-3)node[left]{$v_0$};
		\filldraw	[black,thick]	(5,-3)circle(2pt) -- (7,-3)circle(2pt) --
		(6,-1.25)circle(2pt);
	\end{scope}
	\begin{scope}[shift = {(0.5,0)}]
		\draw [black,thick]	(9,-2)node{$\xrightarrow{\partial}$};
	\end{scope}
	\begin{scope}[shift = {(0,0)}]
		\draw (11,-2)node{$0$};
	\end{scope}
	\begin{scope}[shift = {(-2,0)}]
		\filldraw	[blue!50]	(0,-5.5) -- (1.5,-6) -- (2.5,-4.75) -- (0.5,-4.25);
		\draw	[black,thick]	(0,-5.5) -- (1.5,-6) -- (2.5,-4.75) -- (0.5,-4.25) -- cycle;	\draw	[black,thick]	(0.5,-4.25) -- (1.5,-6)	;	\draw	[black,thick,dashed]	(0,-5			.5) -- (2.5,-4.75);	
		\filldraw	[black,thick]	(0,-5.5)circle(2pt)node[left]{$v_0$} (1.5,-6)circle(2pt)node[below]{$v_1$} (2.5,-4.75)circle(2pt)node[right]{$v_2$} (0.5,-4.25)circle(2pt				)node[above]{$v_3$};
	\end{scope}
	\begin{scope}[shift = {(-1,0)}]
		\draw [black,thick]	(3.5,-5)node{$\xrightarrow{\partial}$};	
	\end{scope} 	
	\begin{scope}[shift = {(0.25,-0.1)}]
		\filldraw	[blue!25, opacity=0.8]	(4.75,-5.75) -- (6.25,-6.25) -- (7.25,-5);
		\draw	[black,thick]	(4.75,-5.75) -- (6.25,-6.25) -- (7.25,-5) -- cycle;
		\filldraw	[black,thick]	(4.75,-5.75)circle(2pt) (6.25,-6.25)circle(2pt) (7.25,-5)circle(2pt);
		
		\filldraw	[blue!25, opacity=0.8]	(4.5,-5.25) -- (7,-4.5) -- (5,-4);
		\draw	[black,thick]	(4.5,-5.25) -- (7,-4.5) -- (5,-4) -- cycle;
		\filldraw	[black,thick]	(4.5,-5.25)circle(2pt) (7,-4.5)circle(2pt) (5,-4)circle(2pt);
		
		\filldraw	[blue!25, opacity=0.8]	(4.5,-5.5) -- (6,-6) -- (5,-4.25);
		\draw	[black,thick]	(4.5,-5.5) -- (6,-6) -- (5,-4.25) -- cycle;
		\filldraw [black,thick] (4.5,-5.5)circle(2pt) (6,-6)circle(2pt) (5,-4.25)circle(2pt);
		
		\filldraw	[blue!25, opacity=0.8]	(6.25,-6) -- (7.25,-4.75) -- (5.25,-4.25);
		\draw	[black,thick]	(6.25,-6) -- (7.25,-4.75) -- (5.25,-4.25) -- cycle;
		\filldraw	[black,thick] (6.25,-6)circle(2pt) (7.25,-4.75)circle(2pt) (5.25,-4.25)circle(2pt);
		\filldraw	[black,thick]	(4.5,-5.5)node[left]{$v_0$};
		\filldraw	[black,thick]	(6.25,-6.25)node[below]{$v_1$};
		\filldraw	[black,thick]	(7.25,-4.75)node[right]{$v_2$};
		\filldraw	[black,thick]	(5,-4)node[above]{$v_3$};
	\end{scope}
	\begin{scope}[shift = {(0.5,0)}]
		\draw [black,thick]	(9,-5)node{$\xrightarrow{\partial}$};
	\end{scope}
	\begin{scope}[shift = {(0,0)}]
		\draw (11,-5)node{$0$};
	\end{scope}
	
	\end{tikzpicture}
	\caption{Boundary operator.}
	\label{fig_boundary}
\end{figure}
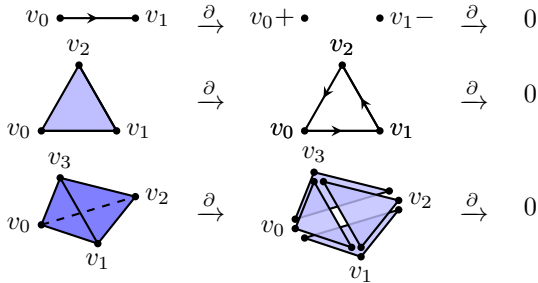

This formula suggests that the boundary of a $k$-simplex is the collection of its $(k-1)$-faces, as illustrated in Figure \ref{fig_boundary}. For example, the boundary of a segment is its two endpoints. A filled triangle is bounded by its three segments. A tetrahedron has its boundary comprised of its four faces which are four triangles. 

The composition of boundary operators gives a chain of complexes:
\begin{equation*}
\cdots \xrightarrow{\partial{}} C_{k+1} \xrightarrow{\partial{}} C_{k} \xrightarrow{\partial{}} C_{k-1} \cdots \xrightarrow{\partial{}} C_{1} \xrightarrow{\partial{}} C_{0} \xrightarrow{\partial{}} 0 
\end{equation*}

Consider two subsets of $C_k(X)$: cycle-subset and boundary-subset, denoted as $Z_k(X)$ and $B_k(X)$ respectively. Let $\ker$ be the kernel and $\ima$ be the image. By definition, we have:
\begin{equation*}
	\begin{array}{l}
		Z_k(X) = \ker(\partial: C_k \to C_{k-1})\\
		B_k(X) = \ima(\partial: C_{k+1} \to C_{k})\\
	\end{array}
\end{equation*}
$Z_k(X)$ includes cycles which are not boundaries while $B_k(X)$ includes only boundaries.  A $k$-cycle $u$ is said to be homologous with a $k$-cycle $v$ if their difference is a $k$-boundary: $[u] \equiv [v] \Longleftrightarrow u - v \in B_k(X)$. A simple computation shows that $\partial \circ \partial = 0$. This result means that a boundary has no boundary. Thus, the $k$-homology of $X$ is the quotient vector space:
\begin{equation*}
H_k(X) = Z_k(X) \backslash B_k(X)
\end{equation*}

The dimension of $H_k(X)$ is called the $k$-th Betti number:
\begin{equation}
\label{eq:bettiComputation}
\beta{}_k = \dim{H_k} = \dim{Z_k} - \dim{B_k}
\end{equation}

This number has an important meaning for coverage problems.  The $k$-th Betti number counts the number of $k$-dimensional holes in a simplicial complex. For example, $\beta{}_0$ counts the connected components while $\beta{}_1$ counts the coverage holes, \etc. \par

\textit{Definition 1 (\v{C}ech complex):} Given $(M,d)$ a  metric space, $\omega$ a finite set of points in $M$ and $\epsilon(\omega)$ a sequence of real positive numbers, the \v{C}ech complex with parameter $\epsilon(\omega)$ of $\omega$, denoted $\check{\mathbf{{C}}}_{\epsilon(\omega)}(\omega)$ is the abstract simplicial complex whose $k$-simplices correspond to non-empty intersection of $(k+1)$ balls of radius $\epsilon(\omega)$ centered at the $(k+1)$ distinct points of $\omega$.\par

If we choose $\epsilon(\omega)$ to be the cell's coverage range $R$, the \v{C}ech complex verifies the exact coverage of the system. In the \v{C}ech complex, each cell is represented by a vertex. A covered space between cells corresponds to a filled triangle, tetrahedron, \etc. In contrast, a coverage hole between cells corresponds to an empty (or non-filled) triangle, rectangle, \etc. \par

\section{Distributed computation of \v{C}ech complex}
\label{sec:dist_computation_Cech_complex}

\subsection{System model}
\label{subsec:system_model}

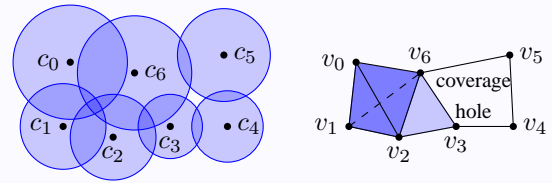
\begin{figure}[width = \textwidth,H!]
\centering
	\begin{tikzpicture}[scale = 0.95]
	\raggedright
	\fill[blue!02, very thick] (-3.5,-0.95)rectangle (5.4,2.2);
		\begin{scope}[shift={(-2,0)},fill opacity = 0.2]
			\fill[blue]	(0,1.05)circle(0.8cm);		
			\fill[blue]	(-0.1,0.15)circle(0.6cm);		
			\fill[blue]	(0.6,0)circle(0.6cm);		
			\fill[blue]	(1.4,0.15)circle(0.45cm);	
			\fill[blue]	(0.9,0.9)circle(0.8cm);		
			\fill[blue] (2.15,1.15)circle(0.65cm);
			\fill[blue] (2.2,0.15)circle(0.5cm);
			\draw[blue]	(0,1.05)circle(0.8cm);		
			\draw[blue]	(-0.1,0.15)circle(0.6cm);		
			\draw[blue]	(0.6,0)circle(0.6cm);		
			\draw[blue]	(1.4,0.15)circle(0.45cm);	
			\draw[blue]	(0.9,0.9)circle(0.8cm);		
			\draw[blue] (2.15,1.15)circle(0.65cm);
			\draw[blue] (2.2,0.15)circle(0.5cm);
			
			\filldraw[black,thick,opacity=1]	(0,1.05)circle(1pt)node[left]{$c_0$};
			\filldraw[black,thick,opacity=1]	(-0.1,0.15)circle(1pt)node[left]{$c_1$};
			\filldraw[black,thick,opacity=1]	(0.6,0)circle(1pt)node[below]{$c_2$};
			\filldraw[black,thick,opacity=1]	(1.4,0.15)circle(1pt)node[right,below]{$c_3$};
			\filldraw[black,thick,opacity=1]	(0.9,0.9)circle(1pt)node[right]{$c_6$};
			\filldraw[black,thick,opacity=1]	(2.15,1.15)circle(1pt)node[right]{$c_5$};
			\filldraw[black,thick,opacity=1]	(2.2,0.15)circle(1pt)node[right]{$c_4$};
			
		\end{scope}
		\begin{scope}[shift={(2,0)},fill opacity = 0.2]
			\filldraw[blue,opacity=0.5]	(0,1.05) -- (-0.1,0.15) -- (0.6,0) -- (0.9,0.9);
			\filldraw[blue,opacity=0.2] (0.6,0) -- (1.4,0.15) -- (0.9,0.9);
			\draw[black,opacity=1] (0,1.05) -- (-0.1,0.15) -- (0.6,0) -- (0.9,0.9) -- cycle;
			\draw[black,opacity=1] (0,1.05) -- (0.6,0);
			\draw[black,opacity=1,dashed] (-0.1,0.15) -- (0.9,0.9);
			\draw[black,opacity=1] (0.6,0) -- (1.4,0.15) -- (0.9,0.9);
			\draw[black,opacity=1] (1.4,0.15) -- (2.2,0.15) -- (2.15,1.15) -- (0.9,0.9);
			\filldraw[black,thick,opacity=1]	(0,1.05)circle(1pt)node[left]{$v_0$};
			\filldraw[black,thick,opacity=1]	(-0.1,0.15)circle(1pt)node[left]{$v_1$};
			\filldraw[black,thick,opacity=1]	(0.6,0)circle(1pt)node[below]{$v_2$};
			\filldraw[black,thick,opacity=1]	(1.4,0.15)circle(1pt)node[below]{$v_3$};
			\filldraw[black,thick,opacity=1]	(0.9,0.9)circle(1pt)node[above]{$v_6$};
			\filldraw[black,thick,opacity=1]	(2.15,1.15)circle(1pt)node[right]{$v_5$};
			\filldraw[black,thick,opacity=1]	(2.2,0.15)circle(1pt)node[right]{$v_4$};
			\draw[black,thick,opacity=1](1.65,0.55)node[align = center]{\footnotesize coverage\\ \footnotesize hole};
		\end{scope}
	\end{tikzpicture}	
\caption{Cells and their \v{C}ech representation.}
\label{fig:fig_Cech}
\end{figure}

We consider a wireless network composed of $N$ distinct cells. We assume that each cell uses isotropic propagation. The coverage of the $i$-th cell is modeled as:
\[
 c_i(v_i, r_i) = \{ x \in \mathbb{R}^2 : \|x - v_i\| \leq r_i \},
\]
where $\|.\|$ is the Euclidean distance, the vertex $v_i$ represents the base station location and $r_i$ is coverage radius of the $i$-th cell. Let $\mathcal{U}$ be the collection of cells, then $\mathcal{U} = \{c_i , i = 0,1,\ldots,(N-1)\}$. The \v{C}ech complex of $\mathcal{U}$, $\check{\mathbf{{C}}}(\mathcal{U})$, is defined as the \v{C}ech complex of the wireless network. In the \v{C}ech complex, each vertex, i.e. a 0-simplex, $v_i$ corresponds to the $i$-th cell $c_i$ in the network. An edge, i.e. a 1-simplex, represents the connection, or the intersection, between two cells. Each $k$-simplex, where  $k \geq 2$, represents the common intersection of the coverage of together $(k+1)$ corresponding cells of this simplex. For example, in the Figure~\ref{fig:fig_Cech}, the 2-simplex $[v_2, v_3, v_6]$ means the overlap of coverage of cell $c_2$, cell $c_3$ and cell $c_6$. There is no coverage hole inside these cells. The higher dimensional simplex is, the higher overlap times is. The 3-simplex $[v_0, v_1, v_2, v_6]$ means the four corresponding cells: $c_0$, $c_1$, $c_2$ and $c_6$, together, have a common intersection. In contrast, a chain of 1-simplices indicates a coverage hole inside corresponding cells of the chain. For example, the chain $[v_3, v_4] + [v_4, v_5] + [v_5, v_6] + [v_6, v_3]$ shows a coverage hole inside four cells $c_3$, $c_4$, $c_5$ and $c_6$. To analyze the network topology, we use characteristics of the homology of the \v{C}ech complex. 

As the definition of the \v{C}ech complex, each $k$-simplex represents the common intersection of together its $(k+1)$ corresponding cells. We denote $\mathbf{S}_k$ the collection of all $k$-simplices of the complex. Each vertex, that is a 0-simplex, corresponds to a cell of the network. The collection of 0-simplices, $\mathbf{S}_0$, is then obviously the list of corresponding vertices of cells.
\[
 \mathbf{S_0} = \{ v_i \mid i = 0, 1, \ldots, (N-1) \}.
\]
Each 1-simplex is an edge that connects two overlapping cells. In other words, it is a pair of two neighbor cells. If two vertices belong to the vertices set of a $k$-simplex, where $k \geq 2$, then they form a 1-simplex. So, they are neighbors. Therefore, each cell needs to detect all its neighbors before computing all its simplices.

\subsection{Neighbors detection}
\label{sec:neighbors_detection}

We assume that each cell $c_i$ can communicate with other cells over radio within a distance $d_i = 2r_i$. We assume that there are enough frequency slots for cells to communicate over radio without collision. Every cell is also connected by a backhaul network. At the initial state, each cell broadcasts a ping message with its position and its radius over radio channel. If a cell receives a ping message, it verifies if the cell that sent this ping message is a neighbor. If they are neighbors, then the cell that received the ping message sends a relationship confirmation together with its position and its radius to the cell that sent the ping message by using the backhaul network. After receiving the confirmation, the cell that sent the ping message adds the cell that sent the confirmation into its collection of neighbors. We assume that all the cells can reply the confirmation within a period $t_{\textnormal{ack}}$. After this period $t_{\textnormal{ack}}$, every cell detects its collection of neighbors. We denote the collection of neighbors of the cell $c_i$ as $\mathbf{N}_i$. 

\subsection{Distributed simplices computation}
\label{subsec:dist_simplices_computation}

When the collection of neighbors is available, each cell computes its simplices by verifying the intersection among it and its neighbors. For the details about the intersection verification, see~\cite{7145759}. As each pair of neighbors forms a 1-simplex, the collection of 1-simplices of each cell is easily found. To find all $k$-simplices, where $k \geq 2$, each cell verifies if a group of it and its $k$ neighbors has common intersection. If these cells have common intersection, then this group forms a $k$-simplex. However, neighborhood is a two-way relationship. Therefore, the verification of intersection could be duplicated by different cells that are neighbors. To avoid the redundant duplication, each cell verifies the intersection by following a right hand rule. This rule is that each cell verifies only with neighbors that are on its right hand side. If there is a neighbor which has the same horizontal coordinate, it verifies only this neighbor if it has higher vertical coordinate. If a simplex is found by a cell, this cell transmits this simplex to every cell that belongs to this simplex. As a result, every cell detects all its simplices. For example, in Figure~\ref{fig:fig_Cech}, the cell $c_2$ verifies the intersection with only the cell $c_3$ and $c_6$. It detects the simplex $[v_2, v_3, v_6]$. It receives its other simplices from the neighbor $c_0$ and $c_1$. The Algorithm~\ref{alg:dist_computation_Cech_complex} reports the distributed computation of \v{C}ech complex for each cell. We denote the cell that is computing the simplices as $c_i$. The highest dimension of simplices that are considered is $\textnormal{dim}_{\textnormal{max}}$. The output of this algorithm, the collection of simplices of the cell $c_i$ is denoted as $\check{\mathbf{C}}_i$. 

\begin{algorithm}[H]
 \caption{Distributed computation of \v{C}ech complex}
 \label{alg:dist_computation_Cech_complex}
 \begin{algorithmic}
  \REQUIRE $c_i$ the cell that is computing; 
  \ENSURE $\check{\mathbf{{C}}}_i$ the collection of simplices of the cell $c_i$;
  \STATE broadcast \{ping, $v_i$, $r_i$\} over radio;
  \STATE $\mathbf{S}_0 = {v_i}$;
  \WHILE {$\textnormal{count time} < t_{\textnormal{ack}}$}
  	\IF {\{ping, $v_j$, $r_j$\} is received from $c_j$}
  	\IF {$d(v_i,v_j) < r_i +r_j$}
  		\STATE add $v_j$ to $\mathbf{N}_i$;
  		\IF {$x_i < x_j$}
  			\STATE add $v_j$ to $\mathbf{S}_0$;
  			\STATE add $[v_i,v_j]$ to $\mathbf{S}_1$;
		\ENDIF
		\IF {$x_i == x_j$ and $y_i <y_j$}
  			\STATE add $v_j$ to $\mathbf{S}_0$;
  			\STATE add $[v_i,v_j]$ to $\mathbf{S}_1$;
		\ENDIF
	\ENDIF
  	\ENDIF
  \ENDWHILE
	\FOR {$k=2$ \textbf{to} $\textnormal{dim}_{\textnormal{max}}$}
  		\FOR {$l=2$ \textbf{to} $C^k_{|\mathbf{S}_0|}$}
	  		\STATE $\hat{s} = l$-th combination of $k$ vertices in $\mathbf{S}_0$;
	  		\STATE $\hat{s} = \hat{s} \cup \{v_i\}$;
	  		\STATE $\hat{c} =$ corresponding cells of $\hat{s}$;
	  		\IF {intersection of all cells in $\hat{c}$ is not empty}
	  			\STATE add $\hat{s}$ to $\mathbf{S}_k$;
	  		\ENDIF
  		\ENDFOR
	\ENDFOR
  \STATE $\check{\mathbf{C}}_i = \{\mathbf{S}_0, \mathbf{S}_1, \ldots, \mathbf{S}_k\}$;
  \STATE send $\check{\mathbf{C}}_i$ to every corresponding cell of vertices in $\mathbf{S}_0$;
  \STATE $n_i = |\mathbf{N}_i|$;
  \WHILE {$n_i > 0$}
	  \IF {$\check{\mathbf{C}}_j$ is received from $c_j$ in $\mathbf{N}_i/\mathbf{S}_0$}
	  	\STATE $\check{\mathbf{C}}_i = \check{\mathbf{C}}_i \cup \check{\mathbf{C}}_j$;
	  	\STATE $n_i = n_i -1$;
	  \ENDIF
  \ENDWHILE
 \RETURN $\check{\mathbf{{C}}}_i$;
 \end{algorithmic}
\end{algorithm}

The global \v{C}ech complex that represents the topology of the whole network is sometimes needed. 
There should be a master cell that controls the topology of the network. This global \v{C}ech complex can be easily built by integrating the simplices computed from every cell. Each cell sends its computed simplices that contain only the vertices satisfied the right hand rule. One more time, this rule is useful, it avoids sending the duplicated simplices.

\section{Applications}
\label{sec:dist_Cech_complex_application}

\subsection{Coverage hole detection}
\label{subsec:coverage_hole_detection}

Coverage hole detection is one fundamental application in solving coverage problem. We introduce a distributed algorithm to detect all the edges that are on the cover (the boundary) of coverage holes. We assume that each cell has computed its simplices as in the Algorithm~\ref{alg:dist_computation_Cech_complex}. If an edge (a 1-simplex) is not a cover edge of a coverage hole, then it musts connect with at least two neighbors that are on two different sides of this edge by filled triangles (a 2-simplex). Otherwise, the edge is a cover edge of a coverage hole. So, each cell cooperates with its neighbors to find all its cover edges. If a pair of neighbors can not find at least two cells that are their common neighbors and are on different sides of the edge formed by this pair, then the edge formed by this pair is a cover edge. 

\begin{algorithm}[H]
 \caption{Distributed coverage hole detection}
 \label{alg:dist_coverage_hole_detection}
 \begin{algorithmic}
  \REQUIRE $c_i(v_i,r_i)$ the cell that is on a cover edge; 
  \ENSURE boundary cycle that contains $v_i$;
  \STATE $m_i = |\mathbf{S}_{1,i}|$;
  \STATE $n_i = |\mathbf{S}_{2,i}|$;
  \STATE $\mathbf{E} = \emptyset$;
  \FOR {$p = 1$ \textbf{to} $m_i$}
  	\STATE $s_1 = p$-th 1-simplex in $\mathbf{S}_{1,i}$;
  	\STATE $\mathbf{V} = \emptyset$;
  	\FOR {}
  		\STATE $s_2 = q$-th 2-simplex in $\mathbf{S}_{2,i}$;
  		\STATE $v = s_2/s_1$;
  		\STATE add $v$ to $\mathbf{V}$;
	\ENDFOR
	\IF {there is no pair of vertices that are on different sides of $s_1$}
	  \STATE add $s_1/v_i$ to $\mathbf{E}$;
  	\ENDIF
  \ENDFOR
  \STATE $h = \{\textnormal{boundary detection}, v_i\}$;
  \STATE send $h$ to every corresponding cell of vertices in $\mathbf{E}$;
  \WHILE {count time $< t_0$}
  	\IF {a boundary detection message $h^\ast$ received}
  		\IF {$v_i$ is in $h^\ast$}
  			\STATE $l = $ vertices list in $h$;
  			\STATE remove vertices in $l$ that are above $v_i$;
  			\STATE add $l$ to the boundary cycle list $\mathbf{L}$;
		\ELSE
			\STATE add $v_i$ to the end of $h^\ast$;
		\ENDIF		
	\ENDIF
  \ENDWHILE
 \RETURN ${\mathbf{{L}}}$;
 \end{algorithmic}
\end{algorithm}

The cells that are on cover edge can cooperate to find the cycle that is boundary of a coverage hole. If a cell $c_i$ is on a cover edge, it sends a boundary detection message \{boundary detection, $v_i$\} to the other cell of this cover edge. If a cell, a receiver, received a boundary detection message from a sender, and it does not correspond to any vertex in the message, it adds its vertex to the end of this message, then it forwards the message to its neighbors that are on cover edges and are not the sender. If this receiver detects that its vertex is on the list of vertices in the message, then it remove all vertices that are above its vertex in the list. It announces the remained vertices as a boundary cycle that covers a coverage hole. We assume that all the message are sent and received in a period $t_0$. After this period, each cell detects all its boundary cycles. The Algorithm~\ref{alg:dist_coverage_hole_detection} describes the details of the coverage hole detection.



\subsection{Coverage optimization}
\label{subsec:coverage_optimization}

Consider a wireless network, we maximize its coverage and minimize its total
transmission power at the same time. Firstly, we ensure a maximal coverage for
the network. Each cell is turned on and is set to work with the highest transmission
power. At this initial state, the network has the largest coverage. However, many cells 
are hardly overlapped. The overlap region between cells causes the waste of 
transmission power due to interference. We can optimize the transmission power 
by minimizing the overlap region. However, the global coverage of the network 
should be conserved. In other words, the two Betti numbers $\beta_0$
and $\beta_1$ of the \v{C}ech complex of the network should be unmodified. The 
optimization problem can be written as:
\begin{equation}
\begin{aligned}
& \underset{r}{\text{min}}
& & \sum_{i=0}^{N-1} r_i^\gamma \\
& \text{s.t.}
& &  \beta_0 = \beta_0^\ast \\
& & &  \beta_1 = \beta_1^\ast \\
& & &  r = (r_0, r_1, \ldots, r_{N-1}), \\
\end{aligned}
\end{equation}
where $\beta_0^\ast$ and $\beta_1^\ast$ are the Betti numbers of the \v{C}ech 
complex of the network at the initial state where every cell is working with the
maximal transmission power.

In this section, we introduce a distributed algorithm to optimize the coverage as well as to save energy for the network. This algorithm is applied for each cell in the network. We assume that all the fenced cells and boundary cells are already known. Only the cells that are not fenced or boundary cells can try to reduce the coverage radius. 

At the first step, each cell needs to search for its neighbors as well as its simplices by following the Algorithm~\ref{alg:dist_computation_Cech_complex}.  
Once the neighbors set is established, each cell now starts its reduction process. 
On each cell, there is a timer which counts down to zero. The timer is set to 
a uniform random value from 1 to $t_{\textnormal{max}}$, where $t_\textnormal{max}$ 
is the maximal value of the timer. 
When the timer of a cell is expired, this cell tries to do a reduction. 
If two cells that are neighbors try to reduce their radius at the same time, 
the coverage hole may not be detected due to the outdated information about 
the radius of each other. Therefore, 
before trying to reduce the radius, each cell sends a ``pause'' message to 
its neighbors. Then, the cell reduces its coverage radius and 
verifies the coverage.
The radius reduction of one cell only makes topology change in the local region that is comprised of
this cell and its neighbors. 
If there is a new coverage hole, it must be inside this local region.
This means that if there is no new
coverage hole after the radius reduction, the Betti numbers $\beta_0$ and $\beta_1$ of the 
\v{C}ech complex of this local region are unchanged.
The verification of the network coverage can be reduced to the coverage verification in only 
this local region as in the Algorithm~\ref{alg:quick_coverage_verification}.

\begin{algorithm}[H]
 \caption{Quick coverage verification after a radius reduction of one cell}
 \label{alg:quick_coverage_verification}
 \begin{algorithmic}
  \REQUIRE \begin{tabular}{l}
	     $c^\ast$ the cell that changed its radius;\\
	     $\mathcal{N} = $ neighbors collection of the cell $c^\ast$;\\
	     $\check{\mathbf{C}}^\ast$ = the \v{C}ech complex of $\{c^\ast\} \cup \mathcal{N}$ \\before the radius reduction;
           \end{tabular}
  \ENSURE \begin{tabular}{l}
           \textbf{true} if and only if there is no new coverage hole.\\
          \end{tabular}
  \STATE compute Betti numbers $\beta_0^\ast$ and $\beta_1^\ast$ of $\check{\mathbf{C}}^\ast$;
  \STATE $\check{\mathbf{C}}$ = the \v{C}ech complex of $\{c^\ast\} \cup \mathcal{N}$ after the radius reduction;
  \STATE compute Betti numbers $\beta_0$ and $\beta_1$ of $\check{\mathbf{C}}$;
  \IF {$\beta_0 = \beta_0^\ast$ and $\beta_1 = \beta_1^\ast$}
    \STATE verification = $\textbf{true}$;
  \ELSE
    \STATE verification = $\textbf{false}$;
  \ENDIF
  \RETURN verification;
 \end{algorithmic}           
\end{algorithm}

If no hole appears, 
the cell confirms the reduction and sends the new value of coverage radius to 
its neighbors. It also sends the ``continue'' message to its neighbors to tell
them that they can continue. If a cell received a ``pause'' message, it pauses 
its process and waits until the message ``continue'' is received. Then,
it continues its process normally.
\begin{algorithm}[h]
 \caption{Distributed downhill energy saving algorithm for each cell}
 \label{alg:distributed_downhill}
 \begin{algorithmic}
  \REQUIRE $c$ a cell in the network;
  \ENSURE the optimal radius for $c$;

  \STATE transmit the position and the coverage radius of $c$ to other cells;
  \STATE collect the information about the position and coverage of other cells;
  \WHILE{(1)}
  	\STATE set timer = uniform$(0,t_{\textnormal{max}})$;
  	\STATE wait until timer expires;
    \STATE $\mathcal{N}_c = $ the collection of neighbors of $c$;
    \IF{a ``pause'' received}
      \STATE wait until ``continue'' received;
    \ENDIF
    \STATE $r_{\textnormal{old}} = r_c$;
    \STATE $r_c = r_c - \Delta{r}_c$;
    \IF {$r_c < r_{c,\textnormal{min}}$}
      \STATE $r_c = 0$;
      \STATE verify the coverage;
      \IF {no coverage hole appears}
	\STATE send ``continue'' to neighbors in $\mathcal{N}_c$;
	\STATE transmit the ``turning off" to other cells;
	\STATE \textbf{break};
      \ELSE
	\STATE $r_c = r_{\textnormal{old}}$;
	\STATE send ``continue'' to neighbors in $\mathcal{N}_c$;
      \ENDIF      
    \ELSE
    \STATE verify the coverage;
      \IF {no coverage hole appears}
	\STATE send $r_c$ to neighbors in $\mathcal{N}_c$;
	\STATE send ``continue'' to neighbors in $\mathcal{N}_c$;
      \ELSE
	\STATE $r_c = r_{\textnormal{old}}$;
	\STATE send ``continue'' to neighbors in $\mathcal{N}_c$;
	\STATE \textbf{break};
      \ENDIF
    \ENDIF
  \ENDWHILE
  \STATE \textbf{return};
 \end{algorithmic}
\end{algorithm} 
There is a special case where two neighbor cells whose timers expire at the 
same time send the ``pause" message to each other simultaneously. One of these 
two cells receives a ``pause" from another before it sends ``continue" 
message to other cells. This cell cancels the current reduction step and sets 
a new value for its timer and waits to retry.

If a cell tries to reduce its coverage radius and makes a coverage hole, 
it reverses its coverage radius to the previous value and stops the reduction 
process. This cell is set to irreducible.

The distributed energy saving algorithm applied for each cell is described in 
the Algorithm~\ref{alg:distributed_downhill}.

\section{Conclusion}
\label{sec:conclusion}

This paper introduces the distributed computation of the \v{C}ech complex aimed for wireless networks. Two distributed applications that are based on this computation of the \v{C}ech complex are also proposed. The first application detects the coverage hole and the second one optimizes the coverage for the network.

\bibliography{biblio}{}
\bibliographystyle{plain}

\end{document}